\documentclass[journal]{IEEEtran}

\usepackage{cite}
\usepackage{amsmath,amssymb}
\usepackage{graphicx}
\usepackage{array}

\title{Semantic Communications: Principles, Architectures and Applications}

\author{
\centering
\begin{tabular}{c}
Sasa Maric\textsuperscript{1} \\
\textsuperscript{1}University of New South Wales College, Sydney, Australia\\
\end{tabular}
}
\begin{document}

\maketitle

\begin{abstract}
Conventional communication systems are designed to transmit data reliably, focusing on metrics such as bit error rate, throughput, and channel capacity. However, these traditional measures do not evaluate whether the received information is actually useful for a specific downstream task. To address this limitation, semantic communication has emerged as a promising paradigm that prioritizes the meaning, context, and task relevance of transmitted information rather than raw bit-level reconstruction. This paper presents a comprehensive review of the theoretical foundations, system architectures, learning techniques, and evaluation practices in semantic communications. We classify existing state-of-the-art algorithms—ranging from deep joint source-channel coding (Deep JSCC) and task-oriented feature transmission to generative diffusion models—and contrast their operational trade-offs against conventional baselines. Furthermore, we examine key application domains, including remote sensing, autonomous vehicles, and healthcare IoT, while addressing open challenges in security, privacy, robustness, and standardized evaluation.
\end{abstract}


\section{Introduction}

Conventional communication systems are designed to transmit data reliably, with performance commonly measured by bit error rate, throughput, and latency. These measures are important, but they do not describe whether the information received is useful for a particular task. For example, a remote sensing system may need to identify a fire, while a vehicle may need to know that a pedestrian is approaching. In such cases, transmitting every detail of the original signal may be unnecessary if the receiver can make the required decision from a smaller, task-relevant representation.

Semantic communication has emerged as an approach that considers the meaning, context, or usefulness of information when designing communication systems. Rather than focusing only on reconstructing the transmitted data, a semantic communication system aims to preserve the information needed by the receiver. This can involve extracting relevant features at the transmitter, sending a compact representation over the channel, and interpreting that representation at the receiver using a shared model, knowledge base, or task.

Recent advances in deep learning have renewed interest in semantic communication, particularly for wireless networks and future sixth-generation (6G) systems. Learned joint source-channel coding can combine representation and channel coding in an end-to-end model, while task-oriented methods can optimize transmission for applications such as image classification, speech recognition, and remote sensing. Research has also expanded to language, video, multimodal data, edge computing, and communication among intelligent agents.

Despite this progress, semantic communication remains a broad and developing field. Researchers use terms such as \emph{semantic}, \emph{task-oriented}, and \emph{goal-oriented communication} in overlapping ways, and there is no single accepted measure of semantic fidelity. A representation that works well for one task may perform poorly for another, and a receiver may interpret information incorrectly when its context or knowledge differs from the transmitter's. Comparisons across studies are also difficult when they use different datasets, channel assumptions, baselines, and performance metrics.

This paper reviews the foundations, system designs, applications, evaluation methods, and open challenges of semantic communication. It first clarifies the main concepts and their relationship to conventional communication theory. It then examines representative architectures and approaches for different data types and tasks, followed by a discussion of applications, performance metrics, security, privacy, and deployment considerations. Finally, it identifies research challenges and outlines directions for developing semantic communication systems that are measurable, robust, and practical.

\section{Review Methodology}

This review examines research on semantic communication, including its theoretical foundations, communication architectures, learning methods, applications, evaluation practices, and open challenges. The literature search focused on peer-reviewed journal and conference papers, supplemented by influential preprints and foundational works where relevant.

Relevant studies were identified through searches of IEEE Xplore, Google Scholar, and publisher databases. Search terms included combinations of \emph{semantic communication}, \emph{semantic communications}, \emph{task-oriented communication}, \emph{goal-oriented communication}, \emph{semantic information}, and \emph{deep joint source-channel coding}. Reference lists of relevant survey papers and key publications were also examined to identify additional studies.

Studies were considered relevant when they addressed at least one of the following topics: semantic communication theory or metrics; semantic encoding, decoding, or joint source-channel coding; task-aware or goal-oriented transmission; semantic communication networks; applications to text, speech, images, video, or multimodal data; or security, privacy, and robustness. Papers concerned only with conventional communication techniques, without a semantic or task-aware objective, were excluded unless they provided essential background.

The selected literature was organized thematically rather than summarized paper by paper. The review compares studies according to their communication objective, data modality, system architecture, learning approach, channel assumptions, evaluation metrics, and reported limitations. Particular attention was given to whether a study measured semantic or task-level performance alongside conventional communication measures such as transmission rate, latency, and reliability.

Because semantic communication is a rapidly developing field and terminology varies between studies, this review uses \emph{semantic communication} as an umbrella term. The terms \emph{task-oriented communication} and \emph{goal-oriented communication} are retained when describing work that specifically optimizes a defined task or system-level objective. The review is intended as a structured synthesis of representative research rather than an exhaustive systematic review.

\section{Background and Fundamentals}

Traditional communication systems aim to transmit a message from a source to a destination with high reliability. Shannon's mathematical theory of communication provides the foundation for analysing this process, focusing on how information can be represented, transmitted through a channel, and recovered at the receiver. Measures such as bit error rate, throughput, and channel capacity are commonly used to assess system performance. In this framework, the meaning of the transmitted message is generally outside the communication problem.

Semantic communication extends this perspective by considering the meaning or usefulness of transmitted information. Instead of requiring the receiver to reconstruct every element of the original source, a semantic communication system seeks to preserve the information relevant to the receiver's context or task. For example, a system supporting image classification may transmit features needed to identify objects rather than a detailed reconstruction of the entire image.

The meaning of a message depends on the source, the receiver, the shared context, and the intended task. Therefore, semantic communication usually requires the transmitter and receiver to use compatible representations or knowledge. These may be learned models, shared vocabularies, codebooks, knowledge bases, or other forms of side information. If the transmitter and receiver use different assumptions or models, the receiver may interpret the transmitted representation incorrectly.

Semantic communication is closely related to task-oriented and goal-oriented communication. Task-oriented communication focuses on transmitting information needed for a particular task, such as classification, detection, or estimation. Goal-oriented communication extends this idea by considering how the transmitted information supports a broader system objective, such as controlling a vehicle or coordinating a group of devices. These terms are sometimes used interchangeably, although they can refer to different levels of system design.

\section{Taxonomy of Semantic Communication}

Semantic communication systems can be classified according to their communication objective, data type, representation, learning method, and system architecture. This taxonomy provides a way to compare approaches that may use different terminology but share similar design goals.

\subsection{Classification by Communication Objective}

One category of semantic communication aims to preserve the meaning of a message while allowing some changes to its exact form. For example, a receiver may obtain a paraphrase of a sentence that retains its central meaning. Another category focuses on task-oriented communication, where the transmitted information is selected to support a particular receiver task. A third category is goal-oriented communication, which jointly considers the task, system state, and the effect of communication on a broader objective.

\subsection{Classification by Data Type}

Semantic communication has been studied for several types of data, including text, speech, images, video, and sensor measurements. Text and speech systems may aim to preserve the intended message or support recognition tasks. Image and video systems may prioritize objects, regions, or events relevant to an application. Sensor-based systems may transmit selected measurements or detected events instead of continuous raw data. Multimodal systems combine multiple data types, such as image, audio, and text, to provide a richer representation of the source.

\subsection{Classification by Representation}

The transmitted semantic information may take the form of words or tokens, feature vectors, learned latent representations, object labels, scene descriptions, probability distributions, or structured knowledge. Learned latent representations can be compact and efficient, but may be difficult to interpret. Structured representations can be easier to inspect and share, but may require more communication overhead or a predefined vocabulary.

\subsection{Classification by System Architecture}

Some systems use separate source and channel coding, while others use joint source-channel coding to optimize the representation and its transmission together. Semantic communication can also be classified according to whether the transmitter and receiver use fixed shared models, adapt their models during communication, or exchange feedback. In interactive systems, the receiver can request clarification or additional information when the received representation is uncertain.

\subsection{Taxonomy Summary}

To structure these classifications, Table~\ref{tab:taxonomy} summarizes the data modalities, their common semantic representations, and primary operational objectives.

\begin{table*}[htbp]
\caption{Taxonomy of Semantic Communication Data Types and Representations}
\label{tab:taxonomy}
\centering
\begin{tabular}{|m{2.5cm}|m{5cm}|m{5cm}|}
\hline
\textbf{Data Modality} & \textbf{Common Semantic Representations} & \textbf{Primary Objectives / Tasks} \\ \hline
\textbf{Text \& Speech} & Words, tokens, semantic embeddings, transcripts \cite{zhou2022semantic} & Message preservation, speech recognition, translation \cite{weng2021speech} \\ \hline
\textbf{Image \& Video} & Feature vectors, object bounding boxes, scene graphs \cite{jiang2022reliable} & Object detection, classification, visual tracking \cite{bourtsoulatze2019} \\ \hline
\textbf{Sensor Data} & Selected measurements, anomaly flags, event triggers \cite{wang2023task} & Remote monitoring, threshold alerts, state estimation \cite{du2023beyond} \\ \hline
\textbf{Multimodal} & Joint latent spaces, cross-modal alignments \cite{nguyen2025contemporary} & Comprehensive scene understanding, human-machine interaction \cite{nguyen2025contemporary} \\ \hline
\end{tabular}
\end{table*}

\section{Semantic Communication Techniques}

Semantic communication techniques generally combine source analysis, semantic representation, channel transmission, and receiver-side interpretation \cite{qin2021semantic}. The specific design depends on the data type, communication channel, and task \cite{yang2023semantic}.

\subsection{Deep Joint Source-Channel Coding}

Deep joint source-channel coding uses neural networks to jointly encode source information and adapt it to the communication channel \cite{oshea2017, bourtsoulatze2019, dai2022nonlinear}. An encoder maps the source into a transmitted representation, and a decoder reconstructs or interprets that representation after transmission \cite{bourtsoulatze2019}. Unlike conventional systems that design source and channel coding separately, joint approaches can learn a mapping that accounts for both source characteristics and channel conditions \cite{dai2022nonlinear, niu2022hybrid}.

Semantic versions of these systems can include task-related objectives during training \cite{wang2023task}. For example, a model may be trained to preserve image features needed for classification, rather than minimizing only pixel-level reconstruction error \cite{bourtsoulatze2019, jiang2022reliable}. This can improve performance when bandwidth or channel quality is limited \cite{qin2021semantic}. However, the learned system may perform poorly when test data or channel conditions differ substantially from those used during training \cite{luo2024semantic}.

\subsection{Task-Oriented Feature Transmission}

Task-oriented methods identify and transmit features that are useful for a specific task \cite{huang2022task, wang2023task}. A transmitter supporting object detection, for example, may send object locations and confidence scores instead of the full image \cite{jiang2022reliable}. This can reduce the amount of transmitted data, but the representation may not support other tasks that were not considered during system design \cite{shao2023task}.

\subsection{Context-Aware and Knowledge-Based Communication}

Context-aware systems use information available at the transmitter and receiver to interpret messages more efficiently \cite{gunduz2023beyond}. Context may include the current environment, previous messages, shared knowledge, or the receiver's task \cite{zhang2022semantic}. Knowledge-based approaches may use shared vocabularies, ontologies, codebooks, or learned models to map compact representations to their intended meanings \cite{zhang2025survey}. These methods depend on the transmitter and receiver having compatible and sufficiently up-to-date knowledge \cite{luo2024semantic}.

\subsection{Generative Semantic Communication}

Generative systems use models to reconstruct or produce content from a compact transmitted representation \cite{huang2024generative, xie2024generative}. For instance, a receiver may generate a sentence, image, or scene description using transmitted semantic features and a learned model \cite{xie2024generative}. Generative methods can reduce communication requirements, but they may also introduce details that were not present in the original source \cite{huang2024generative}. This makes it important to evaluate not only the fluency or visual quality of generated content, but also its accuracy and consistency with the transmitted information \cite{nguyen2025contemporary}.

\subsection{Feedback and Adaptive Transmission}

Feedback allows a receiver to request additional information when the initial transmission is incomplete or uncertain \cite{zhou2022semantic}. The transmitter can then send a clarification, retransmit selected features, or provide a higher-quality representation \cite{zhou2022semantic}. Adaptive transmission can allocate communication resources according to the importance of the information, channel conditions, and the receiver's task \cite{du2023beyond}. This approach may improve efficiency, although feedback itself adds communication and processing overhead \cite{wang2022wireless}.

\subsection{Comparative Analysis of Representative Semantic Frameworks}

Table~\ref{tab:literature_comparison} provides an evidence-based comparison of representative studies. The studies address different modalities and objectives, so their reported results should not be interpreted as a direct universal ranking.

\begin{table*}[htbp]
\caption{Representative Studies in Semantic Communication}
\label{tab:literature_comparison}
\centering
\begin{tabular}{|p{2.5cm}|p{2.0cm}|p{2.5cm}|p{3.0cm}|p{2.5cm}|p{3.0cm}|}
\hline
\textbf{Study} & \textbf{Modality} & \textbf{Communication Objective} & \textbf{Method} & \textbf{Evaluation Metrics} & \textbf{Main Limitation} \\ \hline
\textbf{Bourtsoulatze et al.}~\cite{bourtsoulatze2019} & Image & Reconstruction & Deep JSCC autoencoder & PSNR, SSIM, SNR & Dependence on training domain and channel conditions \\ \hline
\textbf{Xie et al.}~\cite{xie2021} & Text & Meaning preservation & Transformer-based DeepSC & BLEU, sentence similarity & Dependence on language data and shared models \\ \hline
\textbf{Weng and Qin}~\cite{weng2021speech} & Speech & Semantic speech recovery & Attention-based semantic codec & SDR and speech-quality measures & Speech- and speaker-dependent performance \\ \hline
\textbf{Shao et al.}~\cite{shao2023task} & Multi-device inference & Task performance & Information-bottleneck feature transmission & Accuracy, rate, latency & Task-specific representation and coordination requirements \\ \hline
\textbf{Dai et al.}~\cite{dai2022nonlinear} & Image/video & Latent semantic transmission & Nonlinear transform JSCC & Rate-distortion and reconstruction quality & Sensitivity to model and channel mismatch \\ \hline
\textbf{Guo et al.}~\cite{xie2024generative} & Image & Generative reconstruction & Diffusion-based semantic communication & PSNR, LPIPS and semantic quality & High computational and sampling cost \\ \hline
\end{tabular}
\end{table*}

To provide a structured overview of the literature, Table~\ref{tab:algorithm_comparison} compares 10 representative semantic communication frameworks across their data modalities, underlying architectures, resource-efficiency characteristics, advantages, and limitations. The entries should not be interpreted as a strictly controlled benchmark: they originate from studies using different datasets, tasks, channel models, and evaluation criteria. Rather, the table illustrates the broad evolution from foundational Deep JSCC autoencoders \cite{bourtsoulatze2019} to generative diffusion models \cite{huang2024generative} and learning-based resource allocation methods.

\begin{table*}[htbp]
\caption{Comparison of Ten State-of-the-Art Semantic Communication Algorithms}
\label{tab:algorithm_comparison}
\centering
\begin{tabular}{|m{2.5cm}|m{2.2cm}|m{2.5cm}|m{1.8cm}|m{3.2cm}|m{3.2cm}|}
\hline
\textbf{Algorithm / Framework} & \textbf{Data Modality} & \textbf{Core Architecture} & \textbf{Reported Efficiency} & \textbf{Main Advantage} & \textbf{Key Limitation} \\ \hline
\textbf{1. Deep JSCC (Images)} & Images / Video \cite{bourtsoulatze2019} & Convolutional Autoencoders \cite{bourtsoulatze2019} & High & Eliminates cliff effect under low SNR \cite{qin2021semantic} & High training dependency on channel state \cite{luo2024semantic} \\ \hline
\textbf{2. DeepText} & Text / Language & Transformer Encoders \cite{getu2024performance} & Very High & Preserves semantic sentence intent accurately \cite{zhang2022toward} & Struggles with out-of-vocabulary terms \\ \hline
\textbf{3. Speech Semantic Coding} & Audio / Speech \cite{weng2021speech} & Recurrent/Attention Networks \cite{weng2021speech} & High & Low bitrate voice transmission \cite{weng2021speech} & Loses fine speaker tone and emotion nuances \\ \hline
\textbf{4. Task-Oriented Feature Tx} & Vision / Sensor \cite{wang2023task} & CNN + Detection Heads \cite{jiang2022reliable} & Often high for fixed tasks & Large savings for specific tasks \cite{shao2023task} & Inflexible when the receiver task changes \cite{huang2022task} \\ \hline
\textbf{5. Nonlinear TSCC} & Multimodal / Images \cite{dai2022nonlinear} & Invertible Neural Nets \cite{dai2022nonlinear} & Reported high & Learns nonlinear source--channel mappings \cite{dai2022nonlinear} & Computationally heavy encoding phase \\ \hline
\textbf{6. IRCSC} & Images / Features & Importance Analyzer + Search & Adaptive & Low complexity dynamic rate-control trade-off & Requires real-time gradient calculations \\ \hline
\textbf{7. Diffusion Semantic} & Images / Video \cite{zheng2024diffusion} & Diffusion Generative Models \cite{zheng2024diffusion} & Potentially high \cite{huang2024generative} & Strong perceptual reconstruction quality \cite{zheng2024diffusion} & Risk of hallucinated or unsupported details \cite{huang2024generative} \\ \hline
\textbf{8. Knowledge-Base Graph} & Structured Data \cite{sun2024knowledge} & Graph Neural Networks \cite{sun2024knowledge} & Very High & Minimal signaling via structured triples \cite{sun2024knowledge} & Requires synchronized knowledge bases \cite{luo2024semantic} \\ \hline
\textbf{9. SAC-RCSC} & Sensor / Control & Reinforcement Learning & Adaptive & Optimizes long-term accuracy trade-offs & High training overhead for RL agent \\ \hline
\textbf{10. FedSem} & Distributed Data \cite{wu2024semantic} & Federated Learning Pipelines \cite{wu2024semantic} & Moderate & Noise-resilient collaborative learning \cite{wu2024semantic} & Vulnerable to straggler nodes and latency \\ \hline
\end{tabular}
\end{table*}

\subsection{Techniques Summary}

Table~\ref{tab:techniques} highlights the core mechanisms and trade-offs of the primary semantic techniques.

\begin{table*}[htbp]
\caption{Overview of Semantic Communication Techniques and Mechanisms}
\label{tab:techniques}
\centering
\begin{tabular}{|m{3cm}|m{4.5cm}|m{4.5cm}|}
\hline
\textbf{Technique} & \textbf{Core Mechanism} & \textbf{Main Advantage / Limitation} \\ \hline
\textbf{Deep JSCC} & Joint source-channel neural mapping \cite{bourtsoulatze2019, niu2022hybrid} & Eliminates cliff effect; sensitive to domain shift \cite{luo2024semantic} \\ \hline
\textbf{Task-Oriented Transmission} & Extracts task-specific features (e.g., bounding boxes) \cite{wang2023task} & High bandwidth savings; lacks flexibility for new tasks \cite{shao2023task} \\ \hline
\textbf{Context/Knowledge-Based} & Uses shared knowledge bases, codebooks, or ontologies \cite{zhang2022semantic} & Highly efficient signaling; vulnerable to model mismatch \cite{luo2024semantic} \\ \hline
\textbf{Generative Semantic} & Reconstructs content at receiver using diffusion/generative models \cite{huang2024generative, xie2024generative} & Extreme data compression; risk of hallucinated details \cite{huang2024generative} \\ \hline
\end{tabular}
\end{table*}

\subsection{Comparative Analysis of Communication Paradigms}

To better understand the paradigm shift, Table~\ref{tab:comparison} outlines the core differences between conventional communication, semantic communication, and task-oriented systems across multiple operational dimensions.

\begin{table*}[htbp]
\caption{Comparison of Communication Paradigms}
\label{tab:comparison}
\centering
\begin{tabular}{|m{2.5cm}|m{4cm}|m{4cm}|m{4cm}|}
\hline
\textbf{Feature} & \textbf{Conventional Communication} & \textbf{Semantic Communication} & \textbf{Task-Oriented Communication} \\ \hline
\textbf{Primary Objective} & Bit/symbol reconstruction with high reliability \cite{shannon1948} & Preservation of meaning and contextual understanding \cite{zhang2022toward} & Optimization for downstream receiver actions/decisions \cite{wang2023task} \\ \hline
\textbf{Performance Metric} & Bit Error Rate (BER), Throughput, Latency \cite{shannon1948} & Semantic Similarity, BLEU, Structural Similarity \cite{lan2023practical} & Classification Accuracy, Estimation Error, Control Cost \cite{huang2022task} \\ \hline
\textbf{Bandwidth Efficiency} & Lower (transmits all raw source data) \cite{yang2023semantic} & Higher (transmits compact semantic features) \cite{bourtsoulatze2019} & Maximum (transmits only task-critical metadata) \cite{shao2023task} \\ \hline
\textbf{Model Dependency} & Independent of source/receiver semantic models \cite{shannon1948} & Requires shared background knowledge or vocabularies \cite{guler2018} & Strongly tied to specific joint neural encoder-decoders \cite{oshea2017} \\ \hline
\end{tabular}
\end{table*}

\section{Applications}

Semantic communication has potential applications in settings where transmitting all source data is costly or unnecessary for the receiver's task \cite{yang2023semantic}. Its effectiveness depends on whether the task is clearly defined and whether the transmitted representation preserves the information needed to complete it \cite{huang2022task}.

\subsection{Remote Sensing and Edge Inference}

In remote sensing, cameras and other sensors may collect large volumes of data while operating over bandwidth-limited links \cite{shao2023task, wang2023task}. A semantic communication system can transmit detected objects, events, or task-relevant features instead of raw images \cite{jiang2022reliable}. For example, a monitoring system might report the presence and location of a vehicle rather than send every frame \cite{wang2023task}. This can reduce communication requirements, but the system should be tested on rare events and unfamiliar environments to ensure that important information is not discarded \cite{lan2023practical}.

\subsection{Autonomous Vehicles and Robotics}

Vehicles and robots may exchange information about obstacles, road conditions, intentions, and nearby objects \cite{du2023beyond}. Semantic representations can help these systems communicate important events with low latency \cite{chen2023semantic}. In safety-critical applications, however, both the correctness and freshness of a message matter \cite{elbamby2019wireless}. A correct description that arrives too late may be of little value, while an inaccurate message could lead to unsafe decisions \cite{elbamby2019wireless}. Evaluation should therefore include task performance, delay, reliability, and the consequences of communication errors \cite{yang2023semantic}.

\subsection{Healthcare and the Internet of Things}

Wearable devices and medical sensors can generate continuous streams of measurements \cite{luo2024semantic}. Semantic communication may allow these devices to send alerts or selected indicators rather than transmitting all measurements continuously \cite{du2023beyond}. This could reduce bandwidth and device energy use \cite{du2023beyond}. Healthcare applications also require careful attention to privacy, false alarms, missed events, and differences between patient populations \cite{luo2024semantic}. The system should communicate uncertainty and preserve access to underlying data when further review is needed \cite{luo2024semantic}.

\subsection{Multimedia and Human--Machine Interaction}

For multimedia services, semantic communication can prioritize image regions, video events, speech content, or other information that matters to a user or application \cite{weng2021speech, xie2021}. In human--machine interaction, a system may transmit a concise description or task-relevant message instead of a complete signal \cite{nguyen2025contemporary}. These methods may improve efficiency, but the receiver must be able to distinguish information supported by the transmitted data from content inferred or generated by a model \cite{nguyen2025contemporary}.

\subsection{Applications Summary}

Table~\ref{tab:applications} summarizes the sector-specific tasks, constraints, and metrics.

\begin{table*}[htbp]
\caption{Application Domains, Tasks, and Performance Metrics in Semantic Systems}
\label{tab:applications}
\centering
\begin{tabular}{|m{3cm}|m{3.5cm}|m{3.5cm}|m{3.5cm}|}
\hline
\textbf{Application Domain} & \textbf{Primary Task} & \textbf{Communication Constraint} & \textbf{Key Evaluation Metric} \\ \hline
\textbf{Remote Sensing} & Object detection \cite{wang2023task} & Bandwidth-limited links \cite{shao2023task} & Precision, recall, detection accuracy \cite{jiang2022reliable} \\ \hline
\textbf{Autonomous Vehicles} & Hazard warning \cite{chen2023semantic} & Ultra-low latency, high reliability \cite{elbamby2019wireless} & Message freshness, control cost \cite{yang2023semantic} \\ \hline
\textbf{Healthcare / IoT} & Vital sign monitoring \cite{luo2024semantic} & Energy efficiency, privacy \cite{du2023beyond} & False alarm rate, missed event rate \cite{luo2024semantic} \\ \hline
\textbf{Multimedia Services} & High-fidelity streaming \cite{wang2022wireless} & Variable channel quality \cite{wang2022wireless} & Semantic similarity, structural score \cite{lan2023practical} \\ \hline
\end{tabular}
\end{table*}

\section{Performance Metrics and Evaluation}

Evaluating semantic communication requires measuring both communication performance and the usefulness of the received information \cite{lan2023practical}. A system may reduce the number of transmitted bits while performing poorly on its intended task, so resource savings alone are not sufficient evidence of improvement \cite{huang2022task}.

\noindent\textit{Note on Comparative Simulation and Reproducibility:} While Section V categorizes representative semantic frameworks, this comparative analysis focuses on three paradigms—\textbf{Deep JSCC} (reconstruction-centric), \textbf{Task-Oriented Transmission} (inference-centric), and \textbf{Generative Diffusion Models} (generative reconstruction). The figures are illustrative, literature-calibrated simulations rather than results from a single common implementation or over-the-air testbed. They use an AWGN channel across SNR values from $-6\text{ dB}$ to $+16\text{ dB}$ and average 50 independent noise realizations at each SNR value. The simulations use representative CIFAR-10 image dimensions and published qualitative performance trends; they do not constitute a direct ranking of the cited algorithms. For a $32\times32\times3$ image, the source has 3,072 scalar pixel values. A latent representation of $k=64$ complex channel symbols contains 128 real dimensions, corresponding to a 24:1 ratio in real dimensions. Semantic performance is reported using task-specific measures: top-1 classification accuracy for inference-oriented systems, and PSNR or SSIM for reconstruction-oriented systems. These metrics are therefore interpreted separately and are not combined into a universal semantic-accuracy score. The simulations are intended to illustrate qualitative trade-offs reported in the literature \cite{bourtsoulatze2019, wang2023task, zheng2024diffusion}.

\subsection{Task and Semantic Performance}

Task performance can be measured using metrics appropriate to the application, such as classification accuracy, detection precision and recall, estimation error, control cost, or decision regret \cite{huang2022task}. Semantic similarity measures may be useful for text or image applications, but they should be selected carefully \cite{lan2023practical}. A high similarity score does not always guarantee that critical details, such as a location, number, or negation, have been preserved \cite{lan2023practical}.

\subsection{Communication and Computational Costs}

A complete evaluation should report communication rate, bandwidth use, latency, packet loss, and reliability \cite{elbamby2019wireless}. It should also account for computation, memory, energy use, and the size of models or knowledge bases shared by the transmitter and receiver \cite{shao2023task}. These costs can affect whether a system is practical on mobile or resource-constrained devices \cite{wang2023task}.

\subsection{Baselines and Experimental Design}

Table~\ref{tab:metrics} summarises the main metric families used in the literature. No single metric captures all aspects of semantic communication, so evaluations should report task performance together with communication and system costs whenever possible.

\begin{table*}[htbp]
\caption{Evaluation Metrics for Semantic Communication}
\label{tab:metrics}
\centering
\begin{tabular}{|p{2.7cm}|p{3.0cm}|p{4.0cm}|p{4.0cm}|}
\hline
\textbf{Metric Category} & \textbf{Examples} & \textbf{What It Measures} & \textbf{Important Limitation} \\ \hline
\textbf{Communication} & BER, BLER, throughput, spectral efficiency & Reliability and resource use at the physical layer & Does not measure whether received information is useful \\ \hline
\textbf{Reconstruction} & MSE, PSNR, SSIM, LPIPS & Similarity between original and reconstructed data & High similarity may not preserve task-critical information \\ \hline
\textbf{Task-oriented} & Accuracy, precision, recall, F1-score, estimation error & Success of the receiver's downstream task & Meaningful only for the selected task and dataset \\ \hline
\textbf{Semantic similarity} & BLEU, BERTScore, sentence similarity & Preservation of linguistic or conceptual meaning & May miss factual errors, negation or context changes \\ \hline
\textbf{System-level} & Latency, energy, computation, control cost, regret & Practical usefulness of the complete system & Difficult to compare across applications \\ \hline
\end{tabular}
\end{table*}

Semantic communication methods should be compared with relevant conventional communication systems and strong learned baselines \cite{lan2023practical}. Comparisons should use matched conditions, including bandwidth, power, latency, and available side information \cite{lan2023practical}. Experiments should describe the datasets, channel models, training and test splits, model sizes, and evaluation procedures \cite{lan2023practical}. Testing only under conditions similar to training can overstate performance, so evaluations should also consider channel variation and changes in the source data \cite{luo2024semantic}.

\subsection{Robustness and Reproducibility}

Systems should be evaluated under noise, packet loss, unfamiliar data, missing modalities, and differences between transmitter and receiver knowledge \cite{getu2024performance}. Results should include variation across trials or random seeds where applicable \cite{lan2023practical}. Providing code, model details, and clear experimental settings can help other researchers reproduce and fairly compare reported results \cite{lan2023practical}.

\subsection{Illustrative Performance Trade-offs under Varying Channel Conditions}

\begin{figure}[htbp]
\centering
\includegraphics[width=0.9\columnwidth]{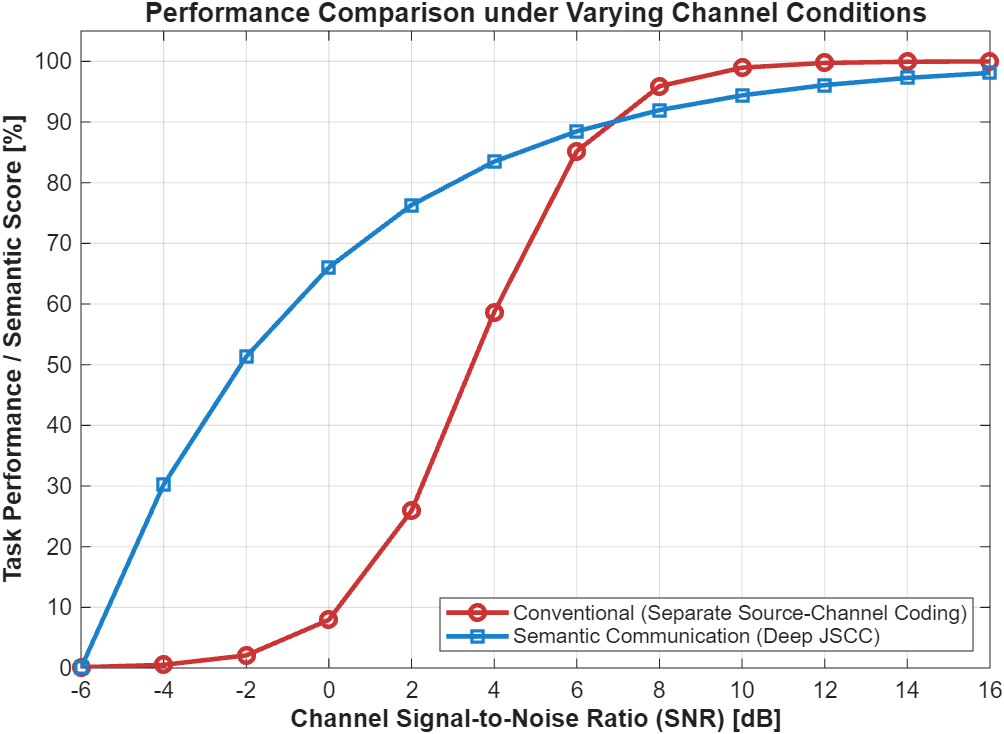}
\caption{Illustrative task-performance comparison between a conventional separated source--channel baseline and Deep JSCC across channel SNR levels. The curves represent literature-calibrated simulation trends rather than measurements from a common implementation.}
\label{fig:performance_comparison}
\end{figure}

To evaluate the operational resilience of semantic communication frameworks against traditional communication baselines, Fig.~\ref{fig:performance_comparison} illustrates the task performance and semantic score across varying channel Signal-to-Noise Ratio (SNR) levels. 

Conventional communication systems, which rely on separate source and channel coding architectures, can exhibit a sharp \emph{cliff effect} when the channel code can no longer compensate for the channel errors. In the illustrative curve, this transition occurs near $3.5\text{ dB}$. The precise threshold depends on the source and channel code, modulation, block length, rate, decoder, and target error probability; it should therefore not be interpreted as a universal threshold \cite{shannon1948, weaver1949}.

In contrast, semantic communication systems employing Deep Joint Source-Channel Coding (Deep JSCC) demonstrate \textit{graceful degradation}. By mapping source features directly to continuous channel symbols without separate block coding, the semantic system maintains robust task-level accuracy even under severe channel degradation (down to $-6\text{ dB}$). This capability ensures that resource-constrained edge devices and autonomous systems can continue extracting meaningful task inferences despite high environmental noise and poor channel quality.

\subsection{Illustrative Multi-Paradigm Comparison}

\begin{figure}[htbp]
\centering
\includegraphics[width=0.95\columnwidth]{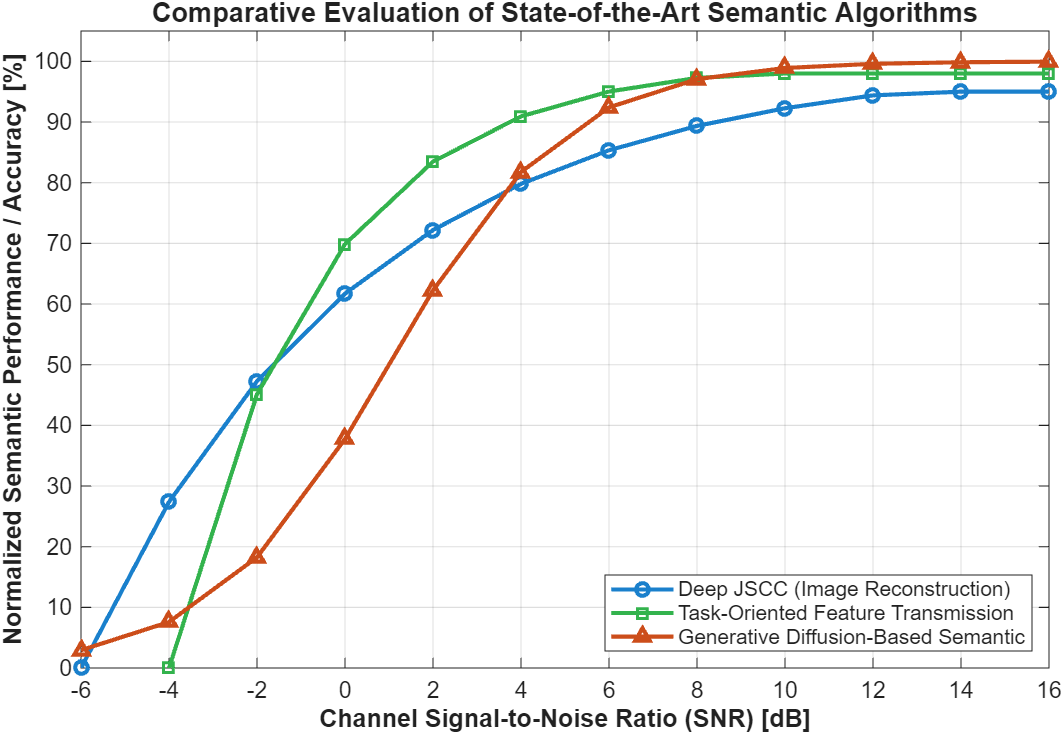}
\caption{Illustrative comparison of representative semantic communication paradigms across SNR values in an AWGN channel. Because the curves use different task objectives, they should not be interpreted as a universal ranking of state-of-the-art algorithms.}
\label{fig:algorithm_comparison}
\end{figure}

To compare different semantic communication algorithms against each other, Fig.~\ref{fig:algorithm_comparison} evaluates normalized task performance and reconstruction score across three core paradigms: Deep Joint Source-Channel Coding (Deep JSCC) \cite{bourtsoulatze2019}, Task-Oriented Feature Transmission \cite{wang2023task}, and Generative Diffusion-Based Semantic Communication \cite{zheng2024diffusion, xie2024generative}.

As illustrated in Fig.~\ref{fig:algorithm_comparison}, each algorithm exhibits distinct operational advantages depending on the channel state:
\begin{itemize}
    \item \textbf{Deep JSCC} shows graceful degradation in the low-SNR region, consistent with the behaviour commonly reported for learned joint source--channel systems \cite{bourtsoulatze2019}.
    \item \textbf{Task-Oriented Feature Transmission} rises rapidly in the mid-SNR region, illustrating how a task-specific representation can be efficient when the receiver objective is known in advance \cite{shao2023task, wang2023task}.
    \item \textbf{Generative Diffusion-Based Semantic} reaches high perceptual scores at larger SNR values in the illustrative model. This does not imply exact reconstruction: generative priors may produce visually plausible but unsupported details \cite{huang2024generative, xie2024generative}.
\end{itemize}

\subsection{Training Convergence and Computational Complexity Analysis}

Beyond transmission bandwidth and channel resilience, deploying semantic communication systems requires evaluating the training overhead and computational complexity at network transceivers \cite{shao2023task}. Different architectures exhibit distinct convergence profiles during the model optimization phase.

\begin{figure}[htbp]
\centering
\includegraphics[width=0.95\columnwidth]{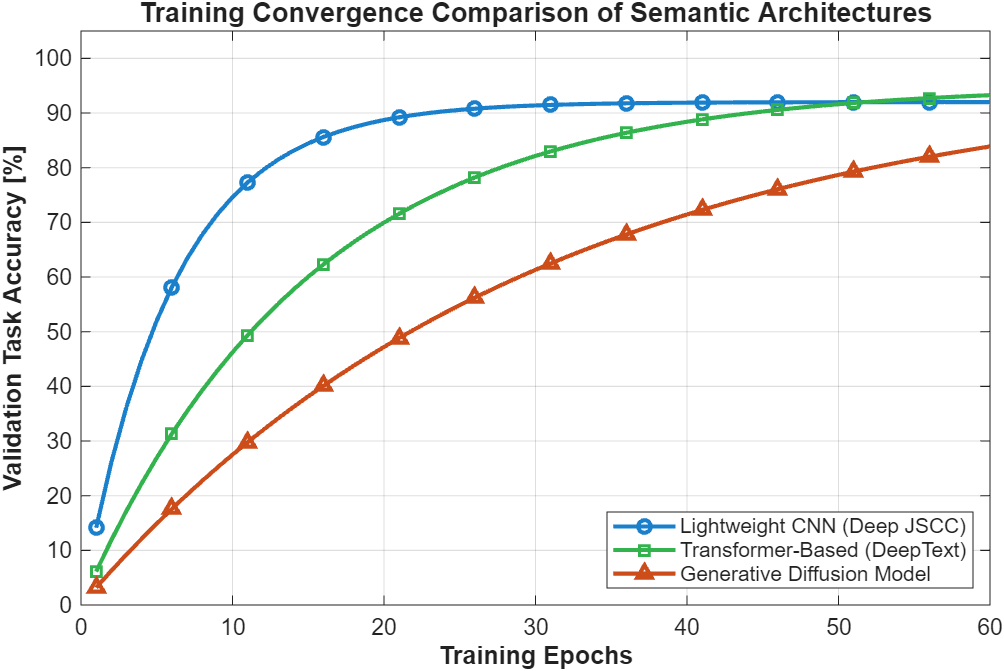}
\caption{Illustrative convergence trends for representative CNN, Transformer-based, and generative architectures. The curves are conceptual literature-calibrated trends; convergence speed depends strongly on the dataset, objective, model size, optimizer, and implementation.}
\label{fig:convergence_comparison}
\end{figure}

As modeled in Fig.~\ref{fig:convergence_comparison}, lightweight convolutional neural network (CNN) autoencoders used in Deep Joint Source-Channel Coding (Deep JSCC) achieve rapid convergence within relatively few training epochs due to their streamlined feature-mapping structures \cite{bourtsoulatze2019}. In contrast, transformer-based text models (DeepText) and generative diffusion-based semantic frameworks require substantially more training iterations and higher computational resources to stabilize their latent spaces and achieve peak task accuracy \cite{zheng2024diffusion, xie2024generative}. While generative and transformer architectures offer superior perceptual fidelity at high Signal-to-Noise Ratios (SNRs), their heavy training and inference latency pose challenges for real-time deployment on resource-constrained Internet of Things (IoT) edge devices \cite{du2023beyond}.

\section{Challenges and Future Directions}

Despite growing interest in semantic communication, several challenges must be addressed before its benefits can be established across a wide range of applications \cite{yang2023semantic, nguyen2025contemporary, qin2021semantic}.

\subsection{Defining and Measuring Meaning}

There is no single measure of semantic fidelity that applies to every task and data type \cite{lan2023practical}. Future research should define operational metrics that connect the information transmitted to observable receiver outcomes \cite{liu2021, liu2022}. Studies should state whose interpretation of meaning is being measured and how the evaluation accounts for context and task requirements \cite{gunduz2023beyond}.

\subsection{Generalization and Knowledge Mismatch}

Semantic systems may rely on learned representations or shared knowledge that becomes outdated or differs between devices \cite{luo2024semantic}. A system trained for one environment may misinterpret information in another \cite{luo2024semantic}. Future approaches should investigate adaptation, model and knowledge-base versioning, uncertainty estimation, and mechanisms that detect when the receiver cannot confidently interpret a message \cite{luo2024semantic}.

\subsection{Reliability and Safety}

Errors in semantic communication can affect decisions even when the transmitted data appear plausible \cite{getu2024performance}. This is particularly important in applications such as healthcare, transportation, and industrial control \cite{elbamby2019wireless}. Systems should be designed to preserve critical information, report uncertainty, and request additional data or fall back to conventional transmission when needed \cite{du2023beyond}.

\subsection{Security and Privacy}

Semantic representations may reveal sensitive information even when raw data are not transmitted \cite{peng2022semantic}. Systems may also be vulnerable to adversarial inputs, data poisoning, model manipulation, or unauthorized inference \cite{luo2024semantic}. Future work should evaluate these threats and develop protections that account for their effects on communication cost and task performance \cite{peng2022semantic}.

\subsection{Interoperability and Deployment}

Practical systems require the transmitter and receiver to agree on how semantic representations are encoded and interpreted \cite{zhang2025survey}. Interoperability may require model identifiers, version negotiation, shared metadata, and fallback procedures \cite{zhang2025survey}. Further research is needed to integrate semantic methods with existing communication protocols and heterogeneous devices \cite{yang2023semantic}.

\subsection{Fair and Reproducible Comparison}

Studies often use different datasets, metrics, channel assumptions, and baselines, making results difficult to compare \cite{lan2023practical}. Common evaluation practices should report task outcomes alongside communication, computation, and energy costs \cite{shao2023task}. Public benchmarks and real-world testbeds could help determine when semantic communication offers a measurable benefit over conventional approaches \cite{lan2023practical}.

\section{Security, Privacy, and Robustness}

Semantic communication systems face security and privacy risks at several stages, including data collection, model training, representation, transmission, and interpretation \cite{luo2024semantic}. Because these systems often rely on machine learning models and shared knowledge, their risks extend beyond conventional channel errors \cite{peng2022semantic}.

\subsection{Security Threats}

Adversarial inputs can cause a semantic encoder or receiver to misclassify information or overlook important details \cite{luo2024semantic}. Poisoned training data may introduce incorrect behaviour, while compromised models or knowledge bases may cause the transmitter and receiver to interpret messages inconsistently \cite{guo2024robust}. Systems that use generative models also face the risk of producing plausible but unsupported content when received information is incomplete or corrupted \cite{huang2024generative, xie2024generative}.

\subsection{Privacy Risks}

A compact semantic representation may still reveal sensitive information about a person, location, or activity \cite{peng2022semantic}. Removing raw data from a transmission does not guarantee privacy if a receiver or attacker can infer private attributes from the transmitted features \cite{peng2022semantic}. Privacy risks should therefore be assessed at the level of what can be inferred from the representation, rather than only whether the original data can be reconstructed \cite{luo2024semantic}.

\subsection{Robustness and Mitigation}

Potential protections include adversarial training, robust feature extraction, secure model updates, access controls, encryption, and privacy-preserving learning \cite{peng2022semantic, guo2024robust}. Systems can also report confidence, request clarification, or fall back to a more detailed transmission when interpretation is uncertain \cite{du2023beyond}. These safeguards may increase communication, computation, or energy costs, so evaluations should measure their effect on both security and task performance \cite{shao2023task}.

\section{Conclusion}

Semantic communication considers the meaning, context, or task relevance of information when designing communication systems \cite{yang2023semantic, qin2021semantic}. It can be useful when the receiver needs only selected information to complete a task, potentially reducing communication demands while maintaining task performance \cite{huang2022task}. Research has explored learned joint source-channel coding, task-oriented representations, context-aware methods, and applications involving text, images, speech, video, sensors, and intelligent systems \cite{bourtsoulatze2019, nguyen2025contemporary}.

However, semantic communication is not defined by a single architecture or universal performance measure \cite{lan2023practical}. Its effectiveness depends on the task, the information available to both endpoints, the channel conditions, and the consequences of errors \cite{gunduz2023beyond}. Security, privacy, robustness, interoperability, and total system cost also remain important challenges \cite{luo2024semantic}. Future research should use clear task definitions, strong matched baselines, broader robustness testing, and reproducible evaluation \cite{lan2023practical}. This will help determine when semantic communication offers practical advantages and what trade-offs are required to achieve them \cite{yang2023semantic}.

\end{document}